\documentclass{article}
\usepackage{amssymb}
\usepackage{amsmath}
\usepackage{amsfonts}
\usepackage{authblk}
\usepackage{enumerate}
\def\Ric{\mathop{\rm Ric}\nolimits}
\newcommand{\hook}{\raisebox{-0.35ex}{\makebox[0.6em][r]
{\scriptsize $-$}}\hspace{-0.15em}\raisebox{0.25ex}
{\makebox[0.4em][l]{\tiny $|$}}}
\begin{document}
\title{Action-angle variables for geodesic motions in 
Sasaki-Einstein spaces $Y^{p,q}$}

\author{Mihai Visinescu\thanks{mvisin@theory.nipne.ro}}

\affil{Department of Theoretical Physics,

National Institute of Physics and Nuclear Engineering,

Magurele, P.O.Box M.G.-6, Romania}

\date{} 
\maketitle

\begin{abstract}

We use the action-angle variables to describe the geodesic motions in the 
$5$-dimensional Sasaki-Einstein spaces $Y^{p,q}$. This formulation allows 
us to study thoroughly the complete integrability of the system. We find that 
the Hamiltonian involves a reduced number of action variables. Therefore one 
of the fundamental frequency is zero indicating a chaotic behavior when the 
system is perturbed.

~


~


~

\end{abstract}

\section{Introduction}

There has been considerable interest recently in Sasaki-Einstein (SE) geometry 
\cite{JS}. In dimension five, an infinite family of explicit SE
metrics  $Y^{p,q}$ on $S^2 \times S^3$ has been constructed, where $p$ and 
$q$ are two coprime positive integers, with $q < p$ \cite{GMSW2}.

A $(2n-1)$-dimensional manifold $M$ is a \emph{contact manifold} if there exists a
$1$-form $\eta$ (called a contact $1$-form) on $M$ such that 
\begin{equation}
\eta \wedge (d \eta)^{n-1} \neq 0\,.
\end{equation}
The \emph{Reeb vector field} $\xi$ dual to $\eta$ satisfies:
\begin{equation}
\eta (\xi) = 1 \quad \text{and} \quad \xi \hook d\eta = 0\,,
\end{equation}
where $\hook$ is the operator dual to the wedge product.

A contact Riemannian manifold $(Y_{2n-1}, g_{Y_{2n-1}})$ is Sasakian if its metric
cone $C(Y_{2n-1}) = Y_{2n-1} \times \mathbb{R}_+$ with the metric
\begin{equation}
ds^2( C(Y_{2n-1})) = dr^2 + r^2 ds^2 (Y_{2n-1})\,,
\end{equation}
is  K\"{a}hler \cite{BG}. Here $r\in (0,\infty)$ may be 
considered as a coordinate on the positive real line $\mathbb{R}_+$. If the 
Sasakian manifold is Einstein, the metric cone is Ricci-flat and K\"{a}hler, 
i.e. Calabi-Yau.

The orbits of the Reeb vector field $\xi$ may or may not close. If the orbits of 
the Reeb vector field $\xi$ are all closed, then $\xi$ integrates to an 
isometric $U(1)$ action on $(Y_{2n-1}, g_{Y_{2n-1}})$. Since $\xi$ is nowhere zero
this action is locally free. If the $U(1)$ action is in fact free, the Sasakian
structure is said to be \emph{regular}. Otherwise it is said to be 
\emph{quasi-regular}. If the orbits of $\xi$ are not all closed, the Sasakian 
structure is said to be \emph{irregular} and the closure of the $1$-parameter
subgroup of the isometry group of $(Y_{2n-1}, g_{Y_{2n-1}})$ is isomorphic to a torus 
$\mathbb{T}^n$ \cite{JS}.

The homogeneous SE metric on $S^2 \times S^3$, known as $T^{1,1}$,
represents an example of regular Sasakian strucure with $SU(2) \times 
SU(2) \times U(1)$ isometry. The $Y^{p,q}$ spaces have isometry $SU(2) \times 
U(1) \times U(1)$ and for $4p^2 - 3 q^2$ a square they are examples of quasi-regular 
SE manifolds. The geometries $Y^{p,q}$ with $4p^2 - 3 q^2$ not a square are
irregular SE spaces.

In a recent paper \cite{BV} the constants of motion for geodesic motions in the 
five-dimensional spaces $Y^{p,q}$ have been explicitly constructed.
This task was achieved using the complete set of Killing vectors and 
Killing-Yano tensors of these toric SE spaces. A multitude of 
constants of motion have been generated, but only five of them are functionally 
independent implying the complete integrability of geodesic flow on $Y^{p,q}$ 
spaces.

The complete integrability of geodesics permits us to construct explicitly the 
action-angle variables. The formulation of an integrable system in these 
variables represents a useful tool for developing perturbation theory.
The action-angle variables define an $n$-dimensional surface which is a 
topological torus (Kolmogorov-Arnold-Moser (KAM) tori) \cite{VIA}.

Our motivation for studying the action-angle parametrization of the phase space 
for geodesic motions in SE spaces comes from recent studies of
non-integrability and chaotic behavior of some classical configuration of strings 
in the context of AdS/CFT correspondence. It was shown that certain classical 
string configurations in $AdS_5 \times T^{1,1}$ \cite{BZ1} or $AdS_5 \times Y^{p,q}$
\cite{BZ2} are chaotic. There were used numerical simulations or an analytic 
approach through the Kovacic's algorithm \cite{JJK}

The purpose of this paper is to describe the geodesic motions in the
SE spaces $Y^{p,q}$ in the action-angle formulation. We
find that the Hamiltonian (energy) involves only four action variables
which have the corresponding frequencies different of zero. One of the
fundamental frequency is zero foreshadowing a chaotic behavior when the 
system is perturbed.

The paper is organized as follows. In the next Section we give the 
necessary preliminaries regarding the metric and the constants of motion for 
geodesics on $Y^{p,q}$ spaces. In Sec. 3 we perform the separation of variables 
and give an action-angle parametrization of the phase space. The paper ends 
with conclusions in Sec. 4.

\section{$Y^{p,q}$ spaces}

The AdS/CFT correspondence represents an important advancement in string theory.
A large class of examples consists of type $IIB$ string theory on the background 
$AdS_5 \times Y_5$ with $Y_5$ a $5$-dimensional SE space. In the frame
of AdS/CFT correspondence $Y^{p,q}$ spaces have played a central role as they
provide an infinite class of dualities.

We write the metric of the 5-dimensional $Y^{p,q}$ spaces \cite{GMSW2,GMSW1,BK} as
\begin{equation}\label{Ypq}
\begin{split}
ds^2_{Y^{p,q}} & = \frac{1-c\, y}{6}( d \theta^2 + \sin^2 \theta\, d \phi^2) 
+  \frac{1}{w(y)q(y)} dy^2 
+ \frac{q(y)}{9} ( d\psi - \cos \theta \, d \phi)^2 \\
& \quad  + 
w(y)\left[ d\alpha + \frac{ac -2y+ c\, y^2}{6(a-y^2)}
(d\psi - \cos\theta \, d\phi)\right]^2\,,
\end{split}
\end{equation}
where
\begin{equation}
w(y)  = \frac{2(a-y^2)}{1-cy} \,, \quad 
q(y)  = \frac{a-3y^2 + 2c y^3}{a-y^2}\,.
\end{equation}
This metric is Einstein with $\Ric g_{Y^{p,q}} = 4 g_{Y^{p,q}}$ 
for all values of the
constants $a,c$. For $c=0$ the metric takes the 
local form of the standard homogeneous metric on $T^{1,1}$ 
\cite{MS}. Otherwise the constant $c$ can be rescaled by a 
diffeomorphism and in what follows we assume $c=1$. 

A detailed analysis of the SE metric $Y^{p,q}$ \cite{GMSW2} showed that for
$0 \leq \theta \leq \pi$ and $0 \leq \phi \leq 2\pi$ the first two terms of 
\eqref{Ypq} give the metric on a round two-sphere. The two-dimensional 
$(y, \psi)$-space defined by fixing $\theta$ and $\phi$ is fibred over this 
two-sphere. The range of $y$ is fixed so that $1-y >0\,,\, a-y^2 >0$ which implies
$w(y) > 0$. Also it is demanded that $q(y) \geq 0$ and that $y$ lies between 
two zeros of $q(y)$, i.e. $y_1 \leq y \leq y_2$ with $q(y_i)=0$.
To be more specific, the roots $y_i$ of the cubic equation
\begin{equation}
a-3y^2 + 2 y^3 = 0 \,,
\end{equation}
are real, one negative $(y_1)$ and two positive, the smallest being $y_2$.
All of these conditions are satisfied if the range of $a$ is
\begin{equation}
0 < a < 1 \,.
\end{equation}
Taking $\psi$ to be periodic with period $2 \pi$, the $(y,\psi)$-fibre at 
fixed $\theta$ and $\phi$ is topologically a two-sphere.
Finally, the period of $\alpha$ is chosen so as to describe a principal $S^1$ 
bundle over $B_4 = S^2 \times S^2$. For any $p$ and $q$ coprime, the space 
$Y^{p,q}$ is topologically $S^2 \times S^3$ and one may take \cite{MS,GMSW2}
\begin{equation}
0 \leq \alpha \leq 2 \pi \ell\,,
\end{equation}
where
\begin{equation}
\ell = \frac{q}{ 3q^2 - 2 p^2 + p(4 p^2 - 3 q^2 )^{1/2}}\,.
\end{equation}

To put the formulas in a simpler forms, in that follows we introduce also 
\begin{equation}
f(y)= \frac{a-2y +y^2}{6(a-y^2)}\,,
\end{equation}
\begin{equation}
p(y)= \frac{w(y) q(y)}{6} =\frac{a-3y^2 + 2y^3}{3(1-y)}\,.
\end{equation}

The conjugate momenta to the coordinates $(\theta,\phi, y, \alpha, \psi)$
are:
\begin{equation}\label{momenta}
\begin{split}
&P_{\theta} =
\frac{1-y}{6} \dot{\theta}\,,\\
&P_y = \frac{1}{6 p(y)} \dot{y}\,,\\
&P_{\alpha}=w(y) \left(\dot{\alpha} + f(y)  \left(\dot{\psi} - \cos\theta
\dot{\phi}\right)\right) \,,\\
&P_{\psi} = w(y) f(y) \dot{\alpha} +
\left[ \frac{q(y)}{9} + w(y) f^2(y)\right]\left(\dot{\psi} - \cos\theta
\dot{\phi}\right)\,,\\
&P_{\phi}  = \frac{1-y}{6} \sin^2\theta \dot{\phi} 
- \cos\theta P_{\psi}\\
&~~~~=  \frac{1-y}{6} \sin^2\theta \dot{\phi} - 
\cos\theta w(y) f(y) \dot{\alpha} 
- \cos\theta\left[\frac{q(y)}{9} +w(y)f^2(y) \right]\dot{\psi}\\
&~~~~~~~+\cos^2\theta\left[ \frac{q(y)}{9} + w(y) f^2(y) \right] \dot{\phi}\,,
\end{split}
\end{equation}
with overdot denoting proper time derivative.

The Hamiltonian describing the motion of a free particle is
\begin{equation}\label{freeHam}
H = \frac12 g^{\mu\nu} P_\mu P_\nu \,,
\end{equation}
which for the $Y^{p,q}$ metric \eqref{Ypq} and using the momenta \eqref{momenta}
has the form:
\begin{equation}\label{HYpq}
\begin{split}
H=&\frac12 \Biggl\{ 6 p(y) P_y^2 + \frac{6}{1-y}\biggl(P_\theta^2 + 
\frac{1}{\sin^2 \theta}(P_\phi  + \cos\theta P_\psi)^2\biggr) + 
\frac{1-y}{2(a-y^2)}P^2_\alpha\Biggr.\\
& \Biggl.+ \frac{9(a-y^2)}{a-3y^2 +2 y^3}\biggl(P_\psi -
\frac{a -2y +y^2}{6(a-y^2)} P_\alpha \biggr)^2\Biggr\}\,.
\end{split}
\end{equation}

Starting with the complete set of Killing vectors and Killing-Yano tensors of the 
SE spaces $Y^{p,q}$ it is possible to find quite a lot of integrals 
of motions \cite{BV,SVV1,SVV2}. However the number of functionally independent 
constants of motion is only five implying the complete integrability of geodesic 
flow on $Y^{p,q}$ spaces. For example we can choose as independent conserved quantities 
the energy
\begin{equation}\label{E}
E=H\,,
\end{equation}
the momenta corresponding to the cyclic coordinates 
$(\phi\,,\, \psi\,,\,\alpha)$
\begin{equation}\label{Pcyc}
\begin{split}
&P_{\phi} = c_{\phi}\,,\\
&P_{\psi} = c_{\psi}\,,\\
&P_{\alpha} = c_{\alpha}\,,
\end{split}
\end{equation}
where $(c_{\phi}\,,\,c_{\psi}\,,c_{\alpha})$ are some constants,
and the total $SU(2)$ angular momentum 
\begin{equation}\label{J2}
\vec{J}^{~2} =P_{\theta}^2 + \frac{1}{\sin^2\theta} \left(P_{\phi}+ 
\cos\theta P_{\psi}\right)^2 + P_{\psi}^2  \,.
\end{equation}

\section{Action-angle variables}

The connection between completely integrable systems and toric geometry in the 
symplectic setting is described by the classical Liouville-Arnold theorem
\cite{VIA,GPS}. 
A dynamical system defined by a given Hamiltonian $H$ on a $2n$-dimensional 
symplectic manifold $(M^{2n},\omega)$ is called Liouville integrable if it admits 
$n$ functionally independent first integrals in involution. In other words, there 
are $n$ functions $\mathbf{F} = (f_1 =H, f_2,\dots,f_n)$  such that
$df_1\wedge \dots \wedge f_n \neq 0$ almost everywhere and
\begin{equation}
\{f_i,f_j\} = 0 \quad, \quad \forall i,j \,.
\end{equation}

Let $\mathbf{F}_\mathbf{c} = (f_1 =E, f_2= c_2,\dots,f_n= c_n)$ by a common 
invariant level set. If $\mathbf{F}_\mathbf{c}$ is regular, compact and connected,
then it is diffeomorphic to the $n$-dimensional Lagrangian torus. 
For $n$ degrees of freedom the motion is confined to an $n$-torus
\begin{equation}\label{it}
\mathbb{T}^n = \underbrace{S^1 \times S^1 \times \cdots \times S^1}_{\text{n~times}} \,.
\end{equation}
These are called \emph{invariant tori} and never intersects taking into account the 
uniqueness of the solution to the dynamical system, expressed as a set of 
coupled ordinary differential equations.

In a neighborhood of $\mathbf{F}_\mathbf{c}$ there are action-angle variables 
$\mathbf{J},\mathbf{w}$ mod $2\pi$, such that the symplectic form becomes
\begin{equation}
\omega = \sum_{i=1}^{n} d J_i \wedge d w_i\,,
\end{equation}
and the Hamiltonian $H$ depends only on actions $J_1,\dots,J_n$. 
An action variable $J_i$ specifies a particular $n$-torus $\mathbb{T}^n$ and is
constant since the tori are invariant. The location on the torus is specified 
by $n$ angle variables $w_i$. Even the system is 
integrable, the dynamics on the singular set (where the differentials of the 
integrals $f_1,\dots,f_n$ are dependent) can be quite complicated \cite{JJD}.

In the case of the geodesic motions on $Y^{p,q}$, for the beginning, we fix a level surface 
$\mathbf{F}=(H, P_\phi,P_\psi,P_\alpha,\vec{J}^{~2})=\mathbf{c}$ of the mutually 
commuting constants of motion \eqref{E}--\eqref{J2}. 
The differentials of the chosen first integrals are real analytic \cite{BV}. 
Then it suffices to require their functional independence at least at one point \cite{BJ}
to apply the Liouville-Arnold theorem.
Further we introduce the generating 
function for the canonical transformation from the coordinates $(\mathbf{p},\mathbf{q})$, 
where $\mathbf{p}$ 
are the conjugate momenta \eqref{momenta} to the coordinates 
$\mathbf{q}=(\theta,\phi,y,\alpha,\psi)$, to the action-angle variables 
$(\mathbf{J},\mathbf{w})$ as the indefinite integral 
\begin{equation}
S(\mathbf{q},\mathbf{c}) = \int_{\mathbf{F}=\mathbf{c}} \mathbf{p}\cdot d\mathbf{q}\,. 
\end{equation}

Since the Hamiltonian \eqref{HYpq} has no explicit time  dependence, we can write
\begin{equation}
S(\mathbf{q},\mathbf{c}) = W(\mathbf{q},\mathbf{c}) - Et\,,
\end{equation}
with the Hamilton's characteristic function 
\begin{equation}\label{Hcf}
W =\sum_i \int p_i d q_i\,. 
\end{equation}

In the case of geodesic motions in SE spaces $Y^{p,q}$ the variables
in the Hamilton-Jacobi equation are separable and consequently we
seek a solution of the Hamilton's characteristic function of the form
\begin{equation}\label{W}
W(y,\theta,\phi, \psi, \alpha)= W_{y}(y) + W_{\theta}(\theta) + W_{\phi}(\phi)
+  W_{\psi}(\psi) +  W_{\alpha}(\alpha)\,.
\end{equation}

The \emph{action variables} $\mathbf{J}$ are defined as integrals over 
complete period of the orbit in the $(p_i,q_i)$ plane
\begin{equation} 
J_i = \oint p_i d q_i = \oint \frac{\partial W_i(q_i;c)}{\partial q_i}
dq_i \qquad \mbox{(no summation)}\,.
\end{equation} 
$J_i$'s form $n$ independent functions of $c_i$'s and can be taken 
as a set of new  constant momenta.

Conjugate \emph{angle variables} $w_i$ are defined by the equations:
\begin{equation}\label{av}
w_i = \frac{\partial W}{\partial J_i} = 
\sum_{j=1}^n \frac{\partial W_j(q_j;J_1,\cdots,J_n)}{\partial J_i}
\end{equation}
having a linear evolution in time
\begin{equation}\label{ff}
w_i = \omega_i t + \beta_i 
\end{equation}
with $\beta_i$ other constants of integration and $\omega_i$ are frequencies
associated with the periodic motion of $q_i$.

Hamilton characteristic functions associated with cyclic variables are
\begin{equation}
\begin{split}
&W_{\phi} = P_{\phi} \phi = c_{\phi} \phi \,,\\
&W_{\psi} = P_{\psi} \psi = c_{\psi} \psi \,,\\
&W_{\alpha} = P_{\alpha} \alpha = c_{\alpha} \alpha \,,
\end{split}
\end{equation}
where $c_\phi,c_\psi,c_\alpha$ are the constants introduced in \eqref{Pcyc}.

The corresponding action variables are 
\begin{equation}
\begin{split}
&J_{\phi} = 2 \pi c_{\phi} \,,\\
&J_{\psi} = 2 \pi c_{\psi} \,,\\
&J_{\alpha} = 2 \pi \ell c_{\alpha}  \,.
\end{split}
\end{equation}

Taking into account \eqref{HYpq} and \eqref{Hcf}, the Hamilton-Jacobi equation becomes
\begin{equation}\label{EJYpq}
\begin{split}
E=&\frac12 \Biggl\{ 6 p(y) \left(\frac{\partial W_y}{\partial y}\right)^2 + 
\frac{6}{1-y}\biggl[\left(\frac{\partial W_\theta}{\partial \theta}\right)^2 + 
\frac{1}{\sin^2 \theta}(c_\phi  + \cos\theta c_\psi)^2\biggr] 
\Biggr.\\
& \Biggl.\frac{1-y}{2(a-y^2)}c^2_\alpha + 
\frac{9(a-y^2)}{a-3y^2 +2 y^3}\biggl[c_\psi -
\frac{a -2y +y^2}{6(a-y^2)} c_\alpha \biggr]^2\Biggr\}\,.
\end{split}
\end{equation}

This equation can be written as follows
\begin{equation}\label{sepJYpq}
\begin{split}
&\left(\frac{\partial W_\theta}{\partial \theta}\right)^2 + 
\frac{1}{\sin^2 \theta}(c_\phi  + \cos\theta _\psi)^2\\
&~~=\frac{1-y}{3}E - p(y)(1-y) \left(\frac{\partial W_y}{\partial y}\right)^2
- \frac{(1-y)^2}{12(a-y^2)}c^2_\alpha\\
&~~~~-\frac{3(a-y^2)(1-y)}{2(a-3y^2 +2 y^3)}\biggl[c_\psi -
\frac{a -2y +y^2}{6(a-y^2)} c_\alpha \biggr]^2\,.
\end{split}
\end{equation}

We observe that the LHS of this equation depends only $\theta$ and independent 
of $y$. Therefore we may set 
\begin{equation}
\left(\frac{\partial W_\theta}{\partial \theta}\right)^2 + 
\frac{1}{\sin^2 \theta}(c_\phi  + \cos\theta c_\psi)^2 = 
c_{\theta}^2\,,
\end{equation}
with $c_{\theta}$ another constant.
From the last equation we can evaluate the action variable 
\begin{equation}\label{J_}
J_{\theta} = \oint d\theta \sqrt{c^2_{\theta} - 
\frac{(c_{\phi} + c_\psi \cos\theta )^2}
{\sin^2\theta}}\,.
\end{equation}

The limits of integrations are defined by the roots $\theta_{-}$ and
$\theta_{+}$ of the expressions in the square root sign and a complete
cycle of $\theta$ involves going from $\theta_{-}$ to $\theta_{+}$
and back to $\theta_{-}$.

This integral can be evaluated by elementary means or using the complex 
integration method of residues which turns out to be more efficient 
\cite{GPS,MV2016}. 
For the evaluation of the integral \eqref{J_} we put $\cos \theta = t$,
extend $t$ to a complex variable $z$ and interpret the integral as a 
closed contour integral in the complex $z$-plane. Consider the integrand
in \eqref{J_}
\begin{equation}\label{integr}
\frac{\sqrt{-(c^2_\theta + c^2_\psi)z^2 - 2c_\phi c_\psi z + c^2_\theta-
c^2_\phi}}{z^2-1} = \frac{\sqrt{-(c^2_\theta + c^2_\psi)}}{z^2-1}
\sqrt{(z-t_+) (z-t_-)}\,,
\end{equation}
where the roots
\begin{equation}
t_{\pm} = \frac{-c_{\phi}c_\psi \pm 
c_{\theta}\sqrt{c^2_{\theta} + 
c^2_\psi -c^2_{\phi}}}{c^2_{\theta} + c^2_\psi}\,,
\end{equation}
are the turning points of the $t$-motion. They are real for
\begin{equation}\label{constr}
c^2_{\theta} + c^2_\psi 
-c^2_{\phi}\geq 0 \,,
\end{equation}
and situated in the interval $(-1,+1)$.

For $z>t_+$ we specify the right side of the square root from \eqref{integr}
as positive. We cut the complex $z$-plane from 
$t_{-}$ to $t_{+}$ and the closed contour integral of the integrand is a loop 
enclosing the cut in a clockwise sense. The contour can be deformed to a large 
circular contour plus two contour integrals about the poles at $z= \pm 1$. 
After simple evaluation of the residues 
and the contribution of the large contour integral we finally get:
\begin{equation}
J_{\theta} = 2\pi\Biggl[\sqrt{c^2_{\theta} + c^2_\psi} - 
c_{\phi} \Biggr] \,.
\end{equation}

For the action variable corresponding to $y$ coordinate we have from 
\eqref{sepJYpq}
\begin{equation}\label{JyYpq}
\begin{split}
\frac{\partial W_y}{\partial y} = & \Biggl\{\frac{1-y}{a - 3y^2 +2 y^3} E - 
\frac{3}{a-3y^2 +2y^3} c_{\theta}^2  \Biggr.\\
& \Biggl. ~- 
\frac{9(a-y^2)(1-y)}{2(a-3y^2+2 y^3)^2} c_{\psi}^2
+ \frac{3(a-2y +y^2)(1-y)}{2 (a-3y^2 +2y^3)^2} c_{\psi}c_{\alpha}
\Biggr.\\
&\Biggl.
~- \frac{(1-y)(2a+a^2 -6ay -2y^2 +2ay^2 +6y^3 -3y^4)}{8(a-3y^2 +2y^3)^2(a-y^2)} 
c_{\alpha}^2
\Biggr\}^{\frac12}\,.
\end{split}
\end{equation}

It is harder to evaluate the action variable $J_y$ in a closed analytic
form taking into account the complicated expression \eqref{JyYpq}. In fact the 
closed-form of $J_y$ is not at all illuminating. More important is the fact that 
$J_y$ depends only of four constants of motion: $E, J_\theta, J_\alpha, J_\psi$. 
In consequence the energy depends only on four action variables 
$J_y, J_\theta, J_\alpha, J_\psi$ representing a reduction of the number of action 
variables entering the expression of the energy of the system.

For the angular variable $w_{\phi}$ we have
\begin{equation}
w_{\phi} = \frac{1}{2\pi} J_{\phi} + \frac{\partial W_\theta}{\partial 
J_\phi}\,.
\end{equation}

Putting $\cos\theta = t$ the second term is
\begin{equation}\label{wphi}
\begin{split}
\frac{\partial W_\theta}{\partial J_\phi}
&=- \frac{1}{2\pi} \int dt\frac{(J_\phi + J_\theta) t^2  + J_\psi t}
{(1-t^2)\sqrt{-(J_\phi + J_\theta)^2 t^2- 2 J_\phi J_\psi t + (J^2_\theta + 
2 J_\theta J_\phi - J^2_\psi)}}\\
&= \frac{1}{2\pi}\int \frac{dt}{1- t^2} \frac{\mathfrak{d} t^2 + \mathfrak{e} t}
{\sqrt{\mathfrak{a} + \mathfrak{b}t + \mathfrak{c}t^2}}\,,
\end{split}
\end{equation}
where
\begin{equation}
\begin{split}
\mathfrak{a}= & J^2_\theta + 2 J_\theta J_\phi - J^2_\psi\,,\\
\mathfrak{b}= & - 2 J_\theta J_\psi \,,\\
\mathfrak{c}= & - (J_\theta + J_\phi)^2\,,\\
\mathfrak{d}= & J_\theta + J_\phi \,,\\
\mathfrak{e}= & J_\psi\,.
\end{split}
\end{equation}

We necessitate the following integrals \cite{GR}:
\begin{equation}
\begin{split}
I_1(\mathfrak{a},\mathfrak{b},\mathfrak{c};t) = 
&\int \frac{dt}{\sqrt{\mathfrak{a} + \mathfrak{b}t + \mathfrak{c}t^2}}\\
=&\frac{-1}{\sqrt{-\mathfrak{c}}}\arcsin 
\Biggl(\frac{2 \mathfrak{c} t + \mathfrak{b}}{\sqrt{-\Delta}}\Biggr)
\end{split}
\end{equation}
evaluated for $\mathfrak{c} <0\,,\,\Delta = 4\mathfrak{a}\mathfrak{c} - \mathfrak{b}^2 <0$,
and
\begin{equation}
\begin{split}
I_2(\mathfrak{a},\mathfrak{b},\mathfrak{c};t) 
+&\int \frac{dt}{t\sqrt{\mathfrak{a} + \mathfrak{b}t + \mathfrak{c}t^2}}\\
=&\frac{1}{\sqrt{-\mathfrak{a}}}\arctan \Biggl(\frac{2\mathfrak{a}+ 
\mathfrak{b}t}{2\sqrt{-\mathfrak{a}}
\sqrt{\mathfrak{a} +\mathfrak{b}t +\mathfrak{c}t^2}}\Biggr)
\end{split}
\end{equation}
evaluated for $\mathfrak{a}<0$. That is the case of the constants 
$\mathfrak{a},\mathfrak{b},\mathfrak{c}$ taking into account the constraint 
\eqref{constr}.

Using these integrals we get for the angular variable $w_\phi$
\begin{equation}
\begin{split}
w_{\phi} =& \frac{1}{2\pi} J_{\phi} -\frac{\mathfrak{d}}{2\pi} 
I_1 (\mathfrak{a},\mathfrak{b},\mathfrak{c};\cos \theta)\\
&- \frac{\mathfrak{d}+\mathfrak{e}}{4 \pi}
I_2(\mathfrak{a}+\mathfrak{b}+\mathfrak{c}, \mathfrak{b} +
2\mathfrak{c},\mathfrak{c};\cos\theta -1)\\
&- \frac{\mathfrak{e}-\mathfrak{d}}{4 \pi}I_2(\mathfrak{a}-\mathfrak{b}+\mathfrak{c}, 
\mathfrak{b}-2\mathfrak{c},\mathfrak{c};\cos\theta +1)\,.
\end{split}
\end{equation}

The explicit evaluation of the angular variables $w_\theta, w_\psi, w_\alpha, w_y$
is again intricate due to the absence of a simple closed-form for the action 
variable $J_y$. However, it is remarkable the fact that one of the fundamental 
frequencies \eqref{ff}
\begin{equation}
\omega_i = \frac{\partial H}{\partial J_i}\,,
\end{equation}
is zero, namely
\begin{equation}\label{fzero}
\omega_\phi = \frac{\partial H}{\partial J_\phi} =0\,,
\end{equation}
since the action $J_\phi$ does not enter the expression of the energy.

The topological nature of the flow of each invariant torus \eqref{it} depends on the 
properties of the frequencies $\omega_i$ \eqref{ff}. There are essentially two cases 
\cite{JP}:
\begin{enumerate}
\item The frequencies $\omega_i$ are nonresonant
\begin{equation}
k_i \omega_i \neq 0 \quad \text{for all} \quad 0\neq k_i \in \mathbb{Z}^n \,.
\end{equation}
Then, on this torus each orbit is dense and the flow is ergodic.
\item The frequencies $\omega_i$ are resonant or rational dependent 
\begin{equation}
k_i \omega_i = 0 \quad \text{for some} \quad 0\neq k_i \in \mathbb{Z}^n \,.
\end{equation}
The prototype is $\mathbf{\omega} = (\omega_1, \cdots , \omega_{n-m}, 0,\cdots,0)$
with $1\leq m\leq n-1$ zero frequencies and $(\omega_1, \cdots , \omega_{n-m})$ 
nonresonant frequencies.
\end{enumerate}

The KAM theorem \cite{VIA} describes how an integrable system reacts to small non-integrable
deformations. The KAM theorem states that for nearly integrable systems, i.e. integrable
systems plus sufficiently small conservative Hamiltonian perturbations, most tori
survive, but suffer a small deformation. However the resonant tori which have rational 
ratios of frequencies get destroyed and motion on them becomes chaotic.

In the case of geodesics on $Y^{p,q}$ space, the frequencies are resonant \eqref{fzero}
giving way to chaotic behavior when the system is perturbed. The analysis performed in 
\cite{BZ2} confirms the present results produced in the action-angle approach.

\section{Conclusions}

The action-angle formulation for $Y^{p,q}$ spaces gives us a better understanding 
of the dynamics of the geodesic motions in these spaces. In spite of the complexity 
of the evaluation of some variables, we are able to prove that the energy of the 
system depends on a reduced number of action variables signaling a degeneracy of
the system.

This fact corroborates a similar result obtained in the case of geodesic motions in 
the homogeneous SE space $T^{1,1}$ \cite{MV2016}. The metric on $T^{1,1}$ may be
written by utilizing the fact that it is a $U(1)$ bundle over $S^2 \times S^2$.
The evaluations of all action and angle variables was completely done putting them
in closed analytic forms. In the case of the space $T^{1,1}$ the isometry is
$SU(2)^2 \times U(1)$ and there are two pairs of fundamental frequencies which are 
resonant. The degeneracy of these two pairs of frequencies may be removed by a
canonical transformation to new action-angle variables. Finally the Hamiltonian 
governing the motions on $T^{1,1}$ can be written in terms of only \emph{three} 
action variables for which the corresponding frequencies are different from zero.

In conclusion, the action-angle approach offers a strong support for the assertion 
that certain classical string configurations in $AdS_5 \times Y_5$ with $Y_5$ in a 
large class of Einstein spaces is non-integrable \cite{BDG,ZE}. It would be 
interesting to extend the action-angle formulations to other five-dimensional SE 
spaces as well as to their higher dimensional generalizations relevant for the 
predictions of the AdS/CFT correspondence.

\section*{Acknowledgements}
The author would like to thank the referee for valuable comments and suggestions
which helped to improve the manuscript.
This work has been partly supported by the project {\it CNCS-UEFISCDI 
PN--II--ID--PCE--2011--3--0137} and partly by the project {\it NUCLEU 16 42 01 
01/2016}.

\end{document}